\documentclass[journal]{IEEEtran}
\usepackage{amssymb}
\usepackage{amsmath} 
\usepackage[utf8]{inputenc}
\usepackage{algorithm}
\usepackage{algorithmic}
\usepackage{float}

\ifCLASSINFOpdf
  \usepackage[pdftex]{graphicx}
  \graphicspath{{./}} 
  \DeclareGraphicsExtensions{.pdf,.jpeg,.png}
\else
  \usepackage[dvips]{graphicx}
  \graphicspath{{./}}
  \DeclareGraphicsExtensions{.eps}
\fi
\ifCLASSINFOpdf
\else
\fi

\usepackage[hyphens]{url}     
\usepackage[breaklinks]{hyperref}

\begin{document}
%
\title{Bridging Large Language Models and Single-Cell Transcriptomics in Dissecting Selective Motor Neuron Vulnerability}
%
%
%

\author{Douglas~Jiang,
        Zilin~Dai, Luxuan Zhang, Qiyi Yu, Haoqi Sun,
        and~Feng~Tian}

%
%




\maketitle

\begin{abstract}
Understanding cell identity and function through single-cell level sequencing data remains a key challenge in computational biology. We present a novel framework that leverages gene-specific textual annotations from the NCBI Gene database to generate biologically contextualized cell embeddings. For each cell in a single-cell RNA sequencing (scRNA-seq) dataset, we rank genes by expression level, retrieve their corresponding NCBI gene descriptions, and transform these descriptions into vector embedding representations using large language models (LLMs). The models used include OpenAI’s text-embedding-ada-002, text-embedding-3-small and text-embedding-3-large (Jan 2024), as well as domain-specific models BioBERT and SciBERT. Embeddings are computed via an expression-weighted average across the top-N most highly expressed genes in each cell, providing a compact, semantically rich representation. This multimodal strategy bridges structured biological data with state-of-the-art language modeling, enabling more interpretable downstream applications such as cell type clustering, cell vulnerability dissection, and trajectory inference.
\end{abstract}


%
\IEEEpeerreviewmaketitle

\section{Introduction}
\IEEEPARstart{S}{elective} vulnerability of neurons is a common feature shared by many neurodegenerative diseases, including Alzheimer's disease (AD), Parkinson's disease (PD), and amyotrophic lateral sclerosis (ALS)~\cite{PANDYA2021101311}. For example, in the case of ALS, a currently incurable motor neuron disease (MND) characterized by muscle loss and paralysis, motor neurons (MNs) are selectively degenerated. Specifically, both upper motor neurons (UMNs) in the motor cortex and lower motor neurons (LMNs) in the spinal cord are affected~\cite{BONANOMI201926, GINSBURG2009277}. The different subtypes of MNs also differ in their vulnerability to ALS-related neurodegeneration~\cite{ROCHAT2015ALS}. $\gamma$ MNs are more resistant to ALS than $\alpha$ MNs. $\alpha$ MNs can be further classified into slow-firing (SF), fast fatigue resistant (FR), and fast fatigable (FF) subtypes, which show an increased vulnerability to ALS~\cite{Blum2021-om}. Importantly, MNs that resist degeneration exhibit significant anatomical and functional differences compared to those that are vulnerable, which provides the key insight in understanding the fundamental neurodegeneration mechanisms in ALS and related neurodegenerative conditions and can be exploited for drug target mining~\cite{RAGAGNIN201913}. 

Advances in high-throughput techniques, such as single-cell RNA sequencing (scRNA-seq), allow researchers to simultaneously quantify the expression levels of numerous genes in a single experiment~\cite{OVENS202122}. However, the resulting datasets can be extremely high-dimensional, containing extensive information across many genes and cell types. It is also very difficult to accurately predict both neuronal identity and their vulnerability in scRNA-seq datasets from neurodegenerative disease mouse models and human patients~\cite {Piwecka2023-zy}. Therefore, predicting neuronal susceptibility and the gene expression pattern underlying it is the unmet need to understand neurodegenerative diseases.

Embedding is an effective and efficient approach for addressing analytical challenges by converting complex, high-dimensional data, such as image, text, or signals, into lower-dimensional numeric representations while maximally preserving essential structural information~\cite{CHANG201821}. Biological datasets typically exhibit high dimensionality and sparsity, which present significant challenges for processing and analysis in bioinformatics. Embedding techniques mitigate these challenges by transforming sequencing data into structured representations but still retaining critical biological insights, thus significantly enhancing the efficacy of bioinformatics analyses. Gene embedding effectively captures functional gene groups and shared annotations, offering significant improvements over traditional differential pathway enrichment analyses typically conducted with RNA-seq data~\cite{SHEININ202522}.

Gene embedding is a method that represents genes as vectors within a high-dimensional space. This method characterizes each gene by considering its expression context, defined by co-expression patterns with other genes. The objective is to optimize these vector representations by maximizing the probability of observing the coexpression context of each gene~\cite{DU201920}.

In this study, we aim to develop and fine-tune an large language model (LLM) capable of analyzing scRNA-seq count matrices to prioritize genes according to their contribution to disease progression. Additionally, we seek to establish an accurate classifier to differentiate neuronal populations based on their susceptibility to various diseases.

\section{Related Work}

\IEEEPARstart{G}{ene} representation and embedding techniques have become essential tools in computational biology, particularly in the analysis of high-dimensional transcriptomic datasets. Early approaches focused on gene coexpression networks, where methods such as Gene2Vec embedded genes based on shared expression contexts between tissues or conditions, effectively capturing functional relationships and improving downstream prediction tasks~\cite{Du2019-ae}. Similarly, embedding strategies utilizing protein-protein interaction networks and Gene Ontology co-annotations have been developed to integrate prior biological knowledge into vector representations~\cite{Edera2022-ha}. While these methods have demonstrated utility in pathway enrichment and disease gene prioritization, they are often limited by fixed biological priors and lack dynamic adaptability to new contexts, such as disease-specific single-cell data.

In parallel, the emergence of LLMs has transformed natural language processing and has begun to influence biological data analysis. Models like BioBERT~\cite{Lee2020-pm} and SciBERT~\cite{Beltagy2019-tr} were pretrained on large biomedical corpora and have demonstrated strong performance in tasks such as named entity recognition, relation extraction, and biomedical question answering. Beyond text mining, recent work has adapted LLMs for biological sequence modeling—such as protein folding prediction with ESM~\cite{Lin2023-zl}, ProteinBERT~\cite{Brandes2022-qz}, and AlphaFold~\cite{Abramson2024-ax}, by treating amino acid sequences as language-like tokens. However, the application of LLMs to structured omics data, such as gene expression matrices, remains underexplored. Few studies have leveraged semantic gene annotations as a modality for interpreting or embedding gene expression.

In the context of scRNA-seq, dimensionality reduction and embedding are critical for visualization, clustering, and classification. Classical methods such as principal component analysis (PCA), t-distributed stochastic neighbor embedding (t-SNE), and Uniform Manifold Approximation and Projection (UMAP) are widely used to uncover cellular heterogeneity~\cite{McInnes2018-mo}. More advanced methods, such as scVI~\cite{Lopez2018-iu} and scPhere~\cite{Ding2021-ay}, incorporate probabilistic modeling or neural architectures to learn latent representations that better preserve biological structure. While effective, these techniques operate solely on numerical gene expression values and do not incorporate textual gene annotations or biological semantics, potentially missing valuable context.

Our work builds on these foundations by introducing a multimodal embedding framework that integrates structured expression data with gene-specific textual descriptions from the NCBI Gene database. By using pretrained LLMs to encode biological descriptions, we bridge the gap between semantic knowledge and quantitative transcriptomic profiles. To our knowledge, this is the first study to systematically evaluate LLM-based gene embeddings in the context of single-cell classification and neuron subtype vulnerability, highlighting both the promise and current limitations of text-derived embeddings in biological settings.

\section{Method}

\subsection{Problem Formulation}

\begin{figure*}[htbp]
    \centering
    \includegraphics[width=0.95\textwidth]{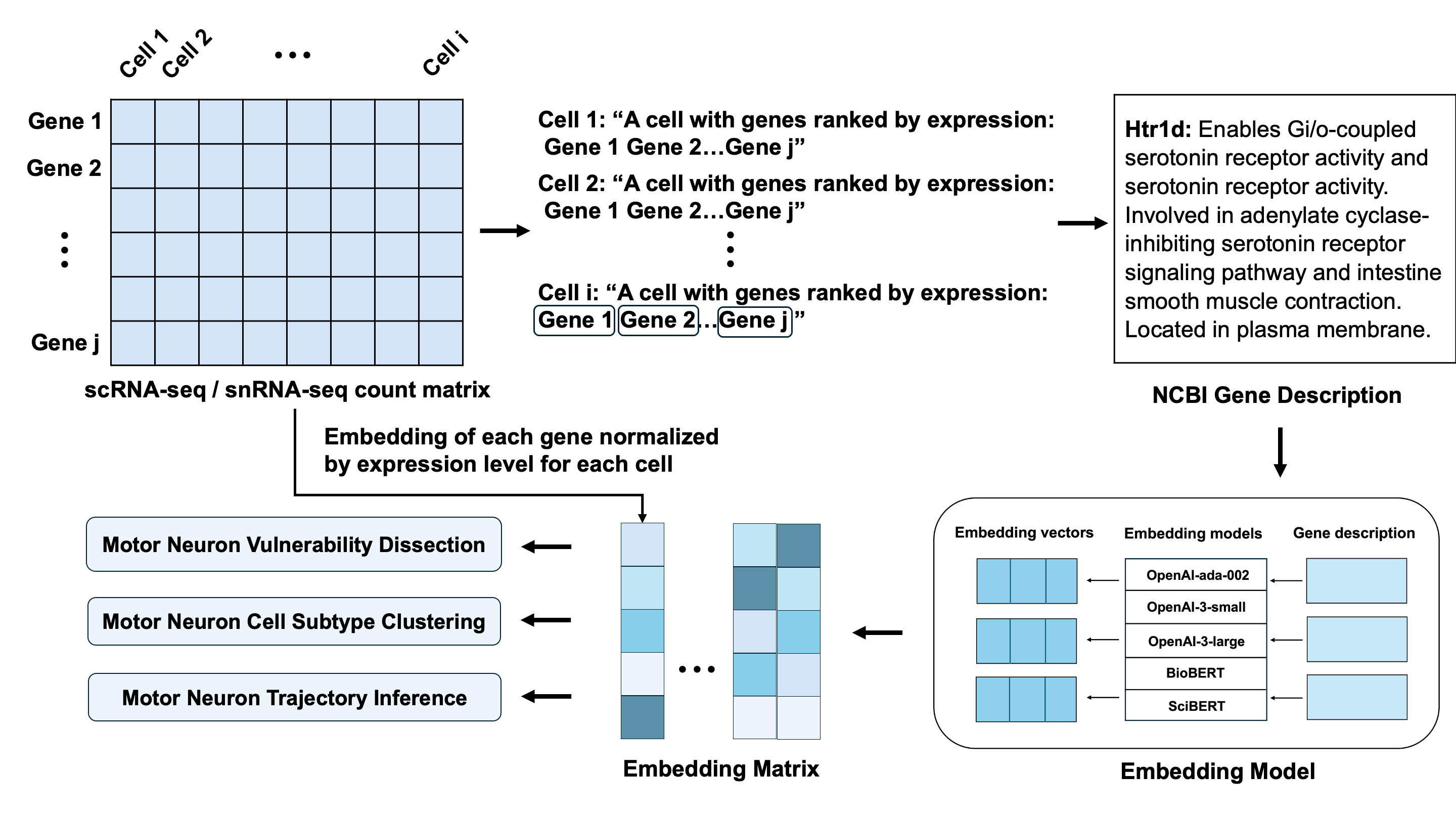}
    \caption{Schematic of the embedding generation pipeline. Gene expression values from a scRNA-seq matrix are used to rank genes per cell. The top ranked genes are mapped to NCBI gene summaries, which are embedded using a language model. Expression-weighted embeddings are averaged to produce cell-level embeddings, which are then used for downstream tasks such as motor neuron vulnerability classification and subtype clustering.}
    \label{fig:pipeline}
\end{figure*}

Consider a scRNA-seq/single-nucleus RNA sequencing (snRNA-seq) dataset represented by an expression matrix \(\mathbf{X}\in\mathbb{R}^{N \times G}\), where \(N\) is the number of cells (MNs) and \(G\) is the number of genes. Each cell \(i\) has an associated subtype label \(y_i\) indicating susceptibility to neurodegenerative diseases such as ALS. Our objective is twofold: (1) prioritize genes relevant to disease progression and (2) develop a robust classifier to differentiate neuronal populations according to their susceptibility.

\subsection{Embedding Generation}
High-dimensional snRNA-seq data are transformed into low-dimensional embeddings using pretrained transformer-based language models. To generate numerical embeddings for snRNA-seq data, we first ranked the genes within each cell by their expression levels, selecting the top expressed genes (e.g., top 100) per cell. For each selected gene, we retrieved its textual biological description from the NCBI Gene database. Each gene description was then individually converted into a numerical embedding vector using transformer-based embedding models. Specifically, the following pretrained embedding models were used independently: OpenAI's text-embedding-ada-002, text-embedding-3-small, text-embedding-3-large, BioBERT, and SciBERT. The embedding vectors generated by each model share the same processing pipeline. For each cell, a final cell-level embedding was calculated by taking a weighted average of the gene-specific embedding vectors, with weights determined by their corresponding gene expression levels. This approach ensures that genes with higher expression contribute proportionally more to the final cell representation. Each gene \(j\) is embedded using:
\begin{equation}
\mathbf{g}_j = \text{Embed}_{\text{LM}}(D_j), \quad j = 1,2,\dots,G,
\end{equation}
where \(D_j\) represents the textual description of gene \(j\), and \(\text{Embed}_{\text{LM}}(\cdot)\) denotes the encoding function of the selected language model.

Cell embeddings $\mathbf{z}_i$ are computed as the weighted average of gene embeddings for the top $N$ expressed genes in each cell. Let $\mathcal{T}_i \subset \{1,\dots,G\}$ denote the set of indices corresponding to the top $N$ expressed genes in cell $i$. Then, the cell embedding is computed as
\[
\mathbf{z}_i = \frac{\sum_{j \in \mathcal{T}_i} x_{ij} \, \mathbf{g}_j}{\sum_{j \in \mathcal{T}_i} x_{ij}}, \quad i = 1, \dots, M,
\]
where $x_{ij}$ represents the expression level of gene $j$ in cell $i$, and $\mathbf{g}_j$ is the embedding vector for gene $j$.

\subsection{Cell Classification Using K-Nearest Neighbors}

To classify MN subtypes with varying susceptibility to neurodegenerative diseases, we implemented a K-Nearest Neighbors (KNN) classifier using the cosine distance metric. Cosine distance, defined as one minus the cosine similarity, captures the angular difference between high-dimensional vectors and is particularly suited for biological data. For two vectors, $\mathbf{x}$ and $\mathbf{v}$, the cosine distance is computed as
\[
d(\mathbf{x}, \mathbf{v}) = 1 - \frac{\mathbf{x} \cdot \mathbf{v}}{\|\mathbf{x}\| \|\mathbf{v}\|}.
\]
In our study, two complementary approaches were employed for classification: one based on pre-computed gene embeddings and the other directly on the raw count matrix. The embedding-based method leverages dimensionality reduction to extract essential features, while the count matrix approach retains the full gene expression information (following appropriate preprocessing).

\paragraph{KNN Classification Using Pre-Computed Embeddings:}  
For a cell represented by its embedding vector $\mathbf{x} \in \mathbb{R}^{d}$, the cosine distance between $\mathbf{x}$ and the $i$th training cell with embedding $\mathbf{E}_i$ is given by

\[
d(\mathbf{x}, \mathbf{E}_i) = 1 - \frac{\mathbf{x} \cdot \mathbf{E}_i}{\|\mathbf{x}\| \|\mathbf{E}_i\|}.
\]

After computing the distances to all training cells, the KNNs are identified. The cell is then classified via a majority vote. The algorithm for this approach is given in Algorithm~\ref{alg:knn_embed}.

\begin{algorithm}[h]
\caption{KNN Classification Using Pre-Computed Embeddings with Cosine Distance}\label{alg:knn_embed}
\begin{algorithmic}[1]
\REQUIRE Embedding matrix $\mathbf{E} \in \mathbb{R}^{N \times d}$, class labels $\{y_i\}_{i=1}^{N}$, query embedding $\mathbf{x} \in \mathbb{R}^{d}$, number of neighbors $k$
\FOR{each training cell $i = 1, \dots, N$}
    \STATE Compute cosine distance:
    \[
    d(\mathbf{x}, \mathbf{E}_i) = 1 - \frac{\mathbf{x} \cdot \mathbf{E}_i}{\|\mathbf{x}\| \|\mathbf{E}_i\|}
    \]
\ENDFOR
\STATE Identify $\mathcal{N}_k(\mathbf{x})$, the indices corresponding to the $k$ smallest distances.
\STATE \textbf{Voting:}
    \begin{itemize}
        \item \textit{Unweighted:} $\hat{y} = \arg\max_{C} \sum_{i \in \mathcal{N}_k(\mathbf{x})} \mathbf{1}\{y_i = C\}$
    \end{itemize}
\STATE \textbf{Output:} Predicted class label $\hat{y}$.
\end{algorithmic}
\end{algorithm}

\paragraph{KNN Classification Using the Direct Count Matrix:}  
In this approach, each cell is represented by its gene expression count vector $\mathbf{x} \in \mathbb{R}^{p}$, where $p$ is the number of genes. After applying appropriate preprocessing (e.g., normalization or log-transformation), the cosine distance between the query cell and the $i$th training cell (with count vector $\mathbf{X}_i$) is computed as
\[
d(\mathbf{x}, \mathbf{X}_i) = 1 - \frac{\mathbf{x} \cdot \mathbf{X}_i}{\|\mathbf{x}\| \|\mathbf{X}_i\|}.
\]
Subsequent neighbor selection and voting steps mirror those of the embedding-based method. The pseudo-code for this approach is outlined in Algorithm~\ref{alg:knn_count}.

\begin{algorithm}[h]
\caption{KNN Classification Using the Direct Count Matrix with Cosine Distance}\label{alg:knn_count}
\begin{algorithmic}[1]
\REQUIRE Count matrix $\mathbf{X} \in \mathbb{R}^{N \times p}$, class labels $\{y_i\}_{i=1}^{N}$, query count vector $\mathbf{x} \in \mathbb{R}^{p}$, number of neighbors $k$
\STATE Preprocess $\mathbf{X}$ (e.g., normalization or log-transformation).
\FOR{each training cell $i = 1, \dots, N$}
    \STATE Compute cosine distance:
    \[
    d(\mathbf{x}, \mathbf{X}_i) = 1 - \frac{\mathbf{x} \cdot \mathbf{X}_i}{\|\mathbf{x}\| \|\mathbf{X}_i\|}
    \]
\ENDFOR
\STATE Identify $\mathcal{N}_k(\mathbf{x})$, the indices corresponding to the $k$ smallest distances.
\STATE \textbf{Voting:}
    \begin{itemize}
        \item \textit{Unweighted:} $\hat{y} = \arg\max_{C} \sum_{i \in \mathcal{N}_k(\mathbf{x})} \mathbf{1}\{y_i = C\}$
    \end{itemize}
\STATE \textbf{Output:} Predicted class label $\hat{y}$.
\end{algorithmic}
\end{algorithm}

This dual approach enables us to evaluate the trade-offs between dimensionality reduction via embedding and the fidelity of MN vulnerability classification using the full gene expression profiles.

\section{Results}

We conducted our analysis using the publicly available dataset \textbf{GSE161621}, which provides snRNA-seq data from the adult mouse spinal cord~\cite{Blum2021-om}. This dataset was generated to characterize by enriching the nuclei of cholinergic neurons (MNs) and non-MN cells from genetically modified reporter mice. Nuclei were isolated via fluorescence-activated nuclei sorting (FANS), the enriched MN and non-MN nuclei were barcoded using 10X Genomics Chromium V3 chemistry and sequenced on both Illumina NextSeq 550 and NovaSeq platforms. In total, the dataset contains the expression profiles for 43,890 nuclei, offering a comprehensive view of the adult mammalian spinal cord.

From this broader population, we focused specifically on MNs based on their marker genes. The resulting expression profiles span multiple MN subtypes, including $\alpha$-FF, $\alpha$-FR, $\alpha$-SF, and $\gamma$ MNs, offering a high-resolution view of transcriptional heterogeneity relevant to MN classification tasks. Specifically, our dataset includes 1,321 $\gamma$ MNs, 986 $\alpha$-FR MNs, 509 $\alpha$-SF MNs, and 492 $\alpha$-FF MNs. These subtypes differ markedly in their susceptibility to degeneration, with $\alpha$-FF MNs being the most vulnerable, followed by $\alpha$-FR and $\alpha$-SF MNs, while $\gamma$ MNs are the most resistant to degeneration~\cite{Blum2021-om}.

We evaluated five different embedding strategies for classifying MN subtypes, comparing their predictive performance using accuracy, weighted F1-score, and Cohen’s kappa (see Table~\ref{tab:performance}). Among all models, text-embedding-ada-002 achieved the best overall performance (accuracy: 70\%, F1: 68\%, kappa: 0.56), followed closely by the domain-specific models BioBERT and SciBERT. In contrast, the newer text-embedding-3 models exhibited reduced performance, with text-embedding-3-small yielding the lowest F1-score (44\%) and Cohen’s kappa (0.21). These results suggest that while general-purpose and domain-specific LLM embeddings capture meaningful biological semantics, not all architectures are equally effective for downstream classification tasks. Model choice, especially when balancing domain-specific pretraining and embedding depth, plays a critical role in the classification of fine-grained neuronal subtypes.

\begin{table}[ht]
\centering
\caption{Comparison of embedding methods for motor neuron classification}
\label{tab:performance}
\begin{tabular}{lccc}
\hline
\textbf{Method} & \textbf{Accuracy} & \textbf{Weighted F1} & \textbf{Cohen's Kappa} \\
\hline
text-embedding-ada-002 & \textbf{0.70} & \textbf{0.68} & \textbf{0.56} \\
text-embedding-3-small & 0.44 & 0.44 & 0.21 \\
text-embedding-3-large & 0.57 & 0.56 & 0.39 \\
BioBERT & 0.66 & 0.66 & 0.51 \\
SciBERT & 0.66 & 0.66 & 0.52 \\
\hline
\end{tabular}
\end{table}

\begin{figure}[htbp]
  \centering
  \includegraphics[width=\linewidth]{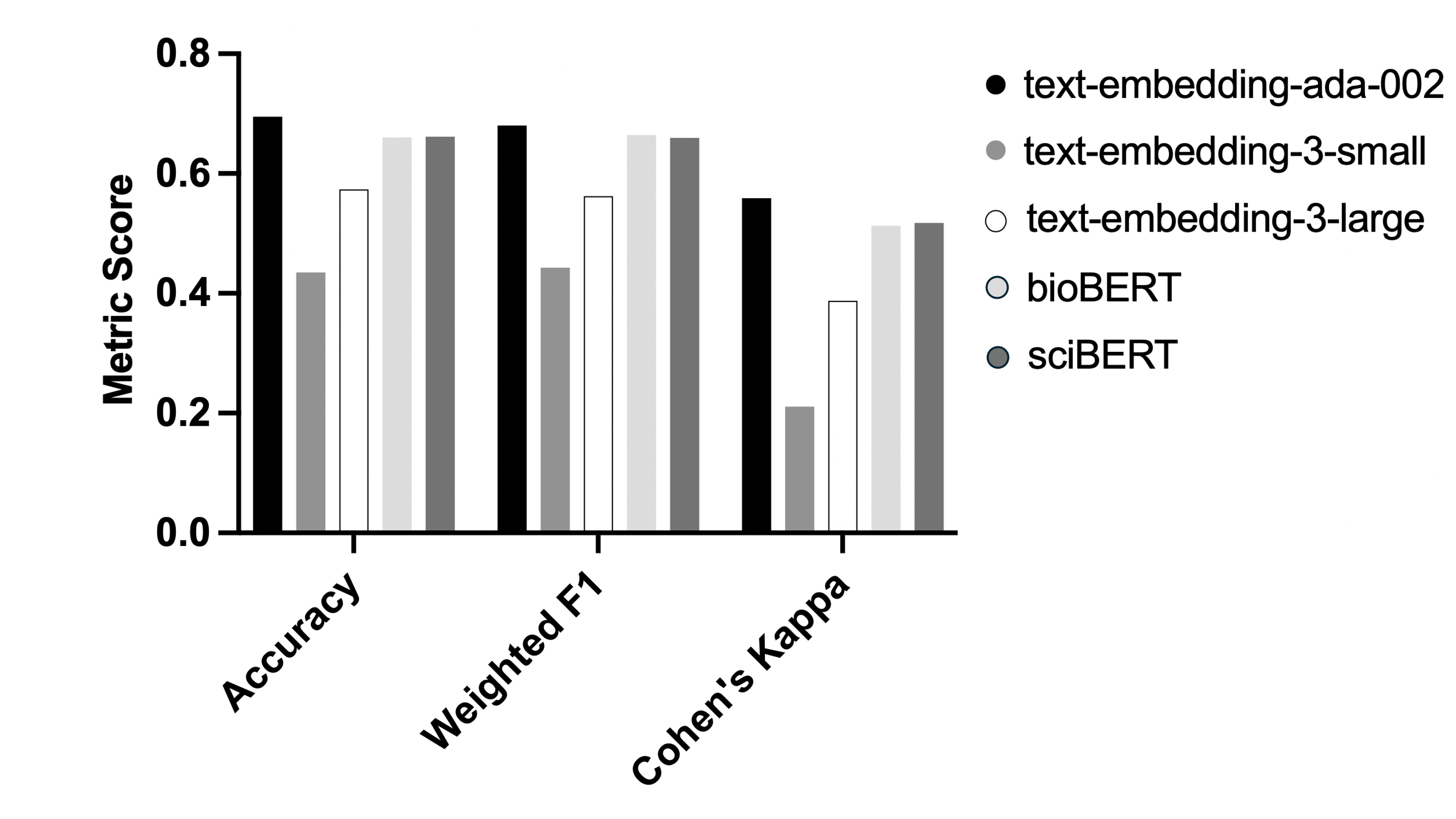}
  \caption{Accuracy, weighted F1, Cohen's Kappa scores for motor neuron subtype classification using different embedding models.}
  \label{fig:accuracy_plot}
\end{figure}

To further investigate how well each embedding model preserved biological structure, we visualized the MN subtypes using UMAP based on their cell-level embeddings and evaluated the clustering quality using three standard metrics: Silhouette score, Calinski–Harabasz index, and Davies–Bouldin score (Table~\ref{tab:UMAP performance}). However, we note that these clustering metrics are known to be less reliable in high-dimensional spaces due to the curse of dimensionality, where pairwise distances become increasingly uniform, density estimates fluctuate, and cluster boundaries blur. Therefore, these metrics should be interpreted with caution in the context of LLM-derived embeddings. Among all models, text-embedding-ada-002 exhibited the most distinct separation visually, especially for \(\gamma\) MNs. However, in terms of clustering metrics, BioBERT achieved the highest Silhouette score (0.0018), indicating better compactness and separation of clusters, while ada-002 led in Calinski–Harabasz score (62.22), suggesting strong between-cluster dispersion. SciBERT also performed competitively, especially in the Davies–Bouldin score (11.79), reflecting lower intra-cluster variance. In contrast, text-embedding-3-small and text-embedding-3-large showed weaker clustering both qualitatively and quantitatively, with more overlapping subtype distributions and worse scores across all metrics. These findings further support the conclusion that domain-specific models such as BioBERT and SciBERT better capture transcriptomic structure than general-purpose models, with ada-002 providing the most balanced performance overall.

\begin{figure*}[htbp]
  \centering
  \includegraphics[width=\textwidth]{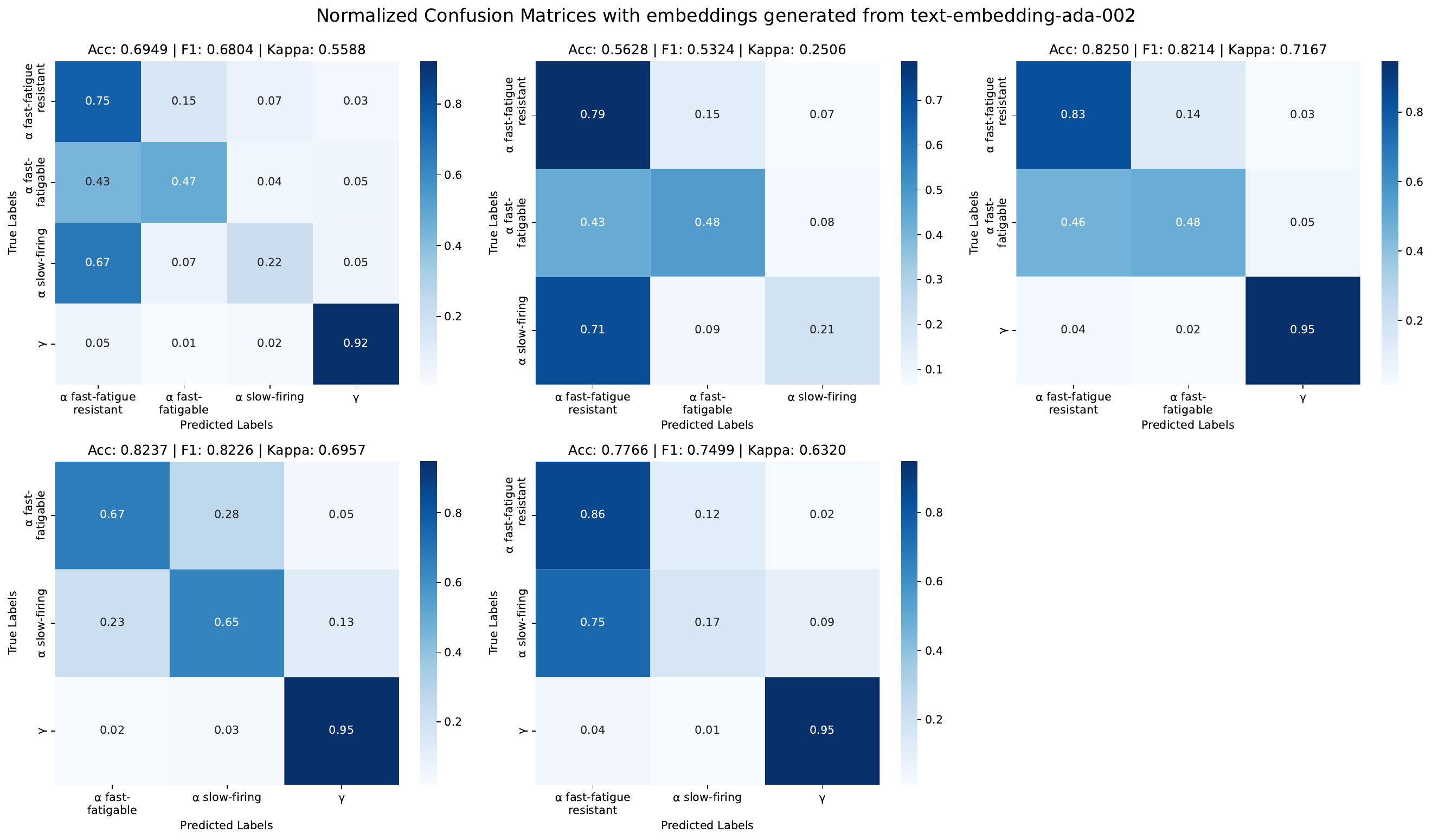}
  \caption{Confusion matrix using text-embedding-ada-002 embeddings.}
  \label{fig:cm_ada002}
\end{figure*}

\begin{figure*}[htbp]
  \centering
  \includegraphics[width=\textwidth]{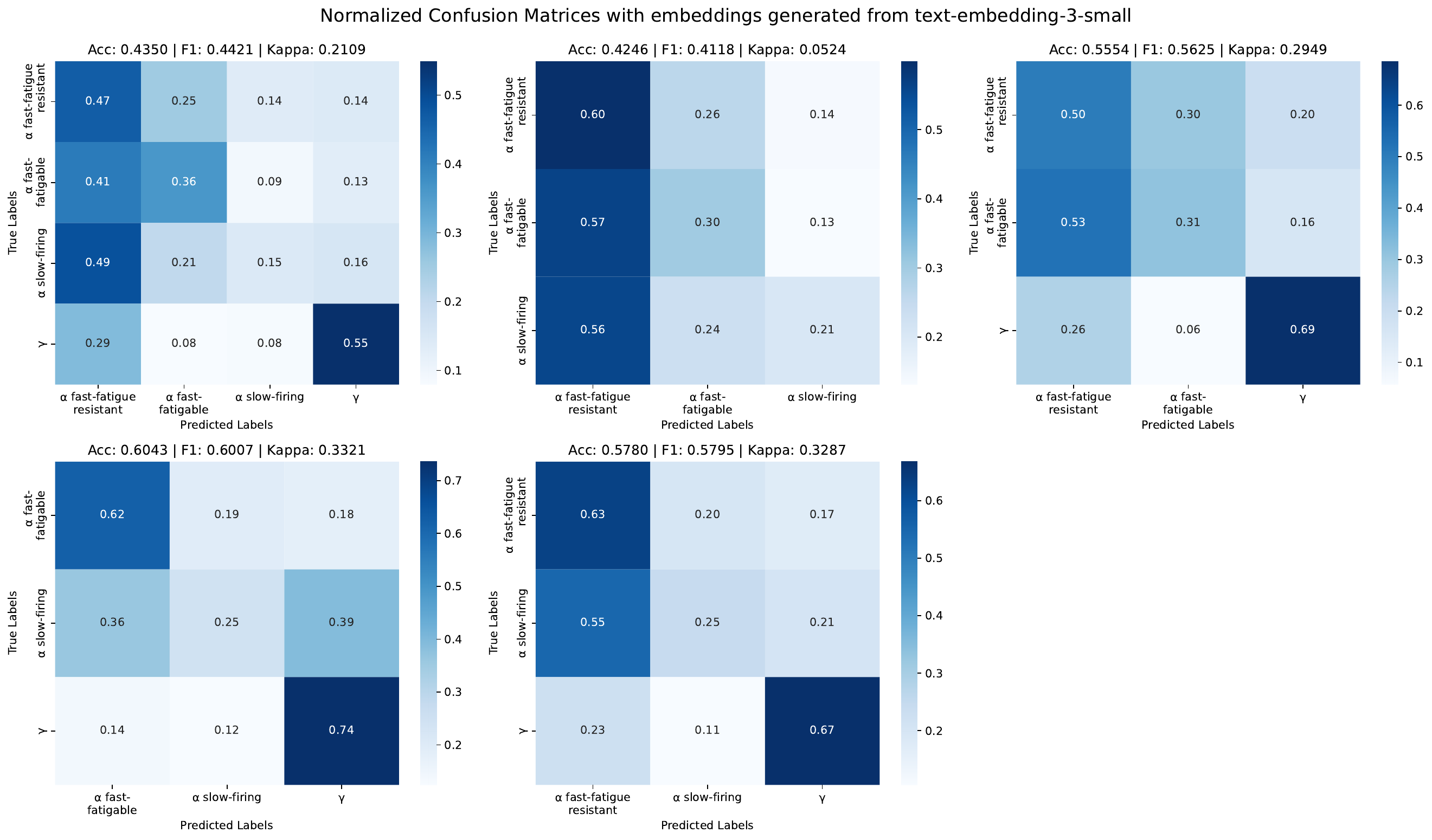}
  \caption{Confusion matrix using text-embedding-3-small embeddings.}
  \label{fig:cm_3small}
\end{figure*}

\begin{figure*}[htbp]
  \centering
  \includegraphics[width=\textwidth]{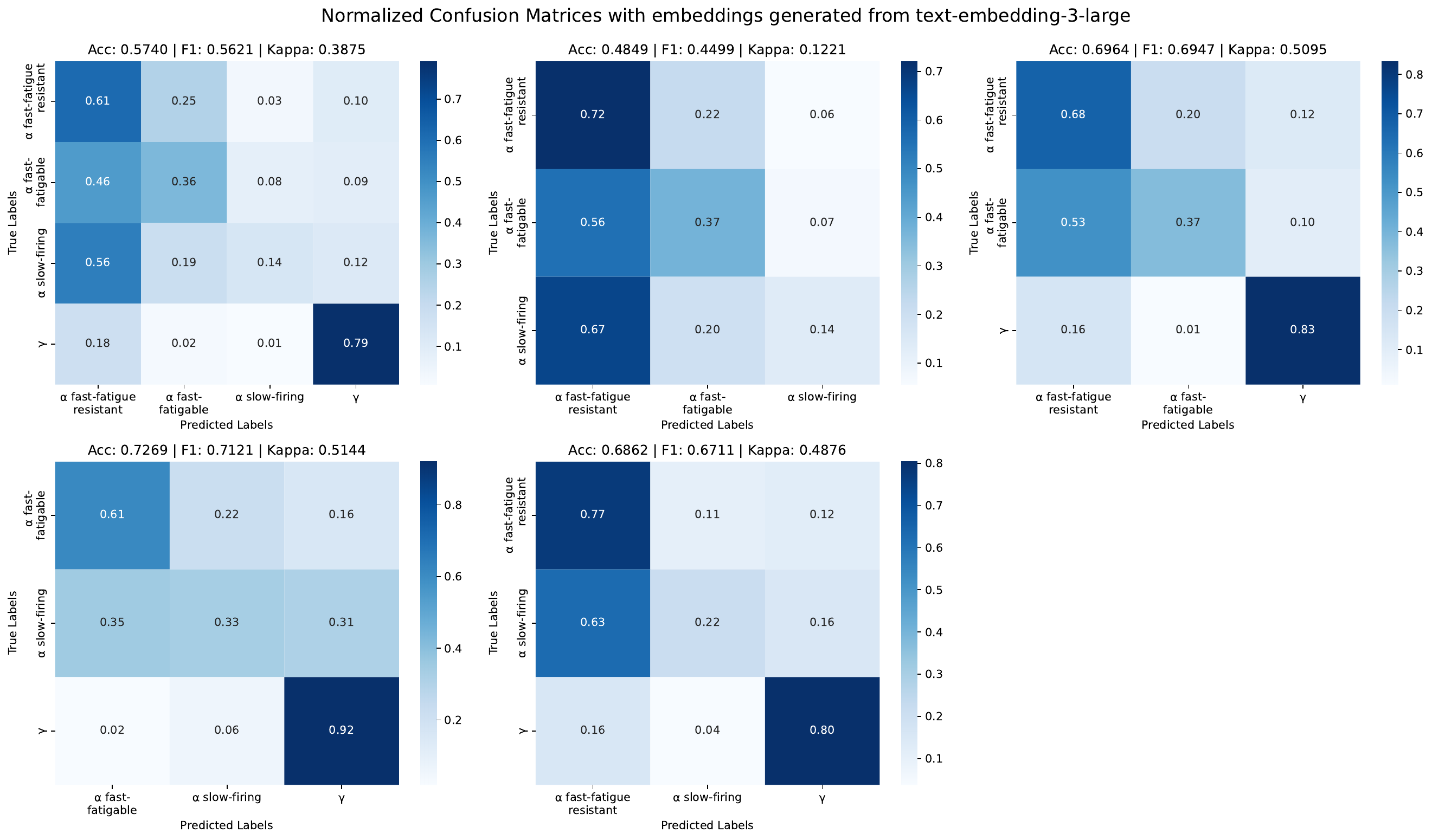}
  \caption{Confusion matrix using text-embedding-3-large embeddings.}
  \label{fig:cm_3large}
\end{figure*}

\begin{figure*}[htbp]
  \centering
  \includegraphics[width=\textwidth]{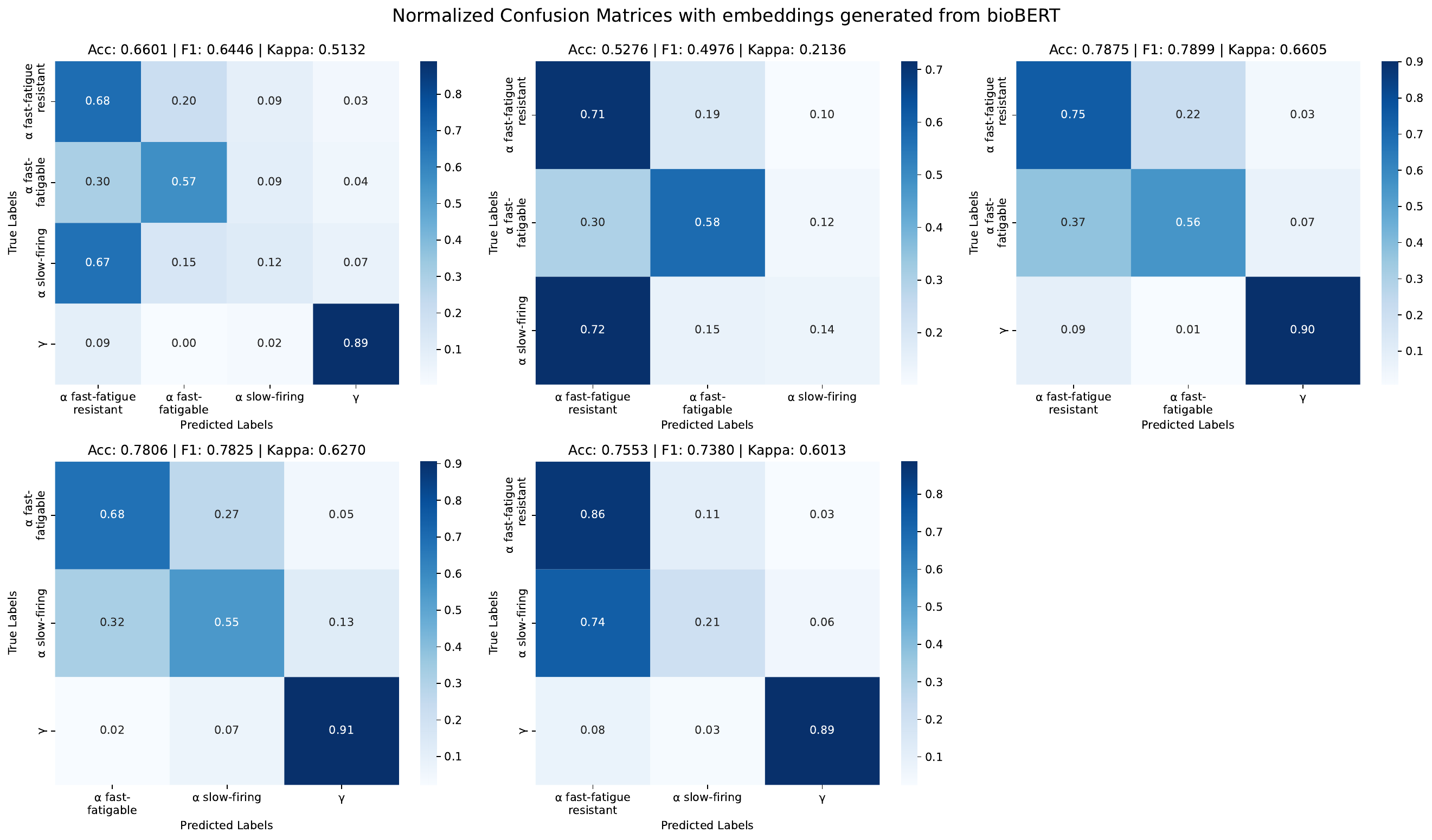}
  \caption{Confusion matrix using BioBERT embeddings.}
  \label{fig:cm_biobert}
\end{figure*}

\begin{figure*}[htbp]
  \centering
  \includegraphics[width=\textwidth]{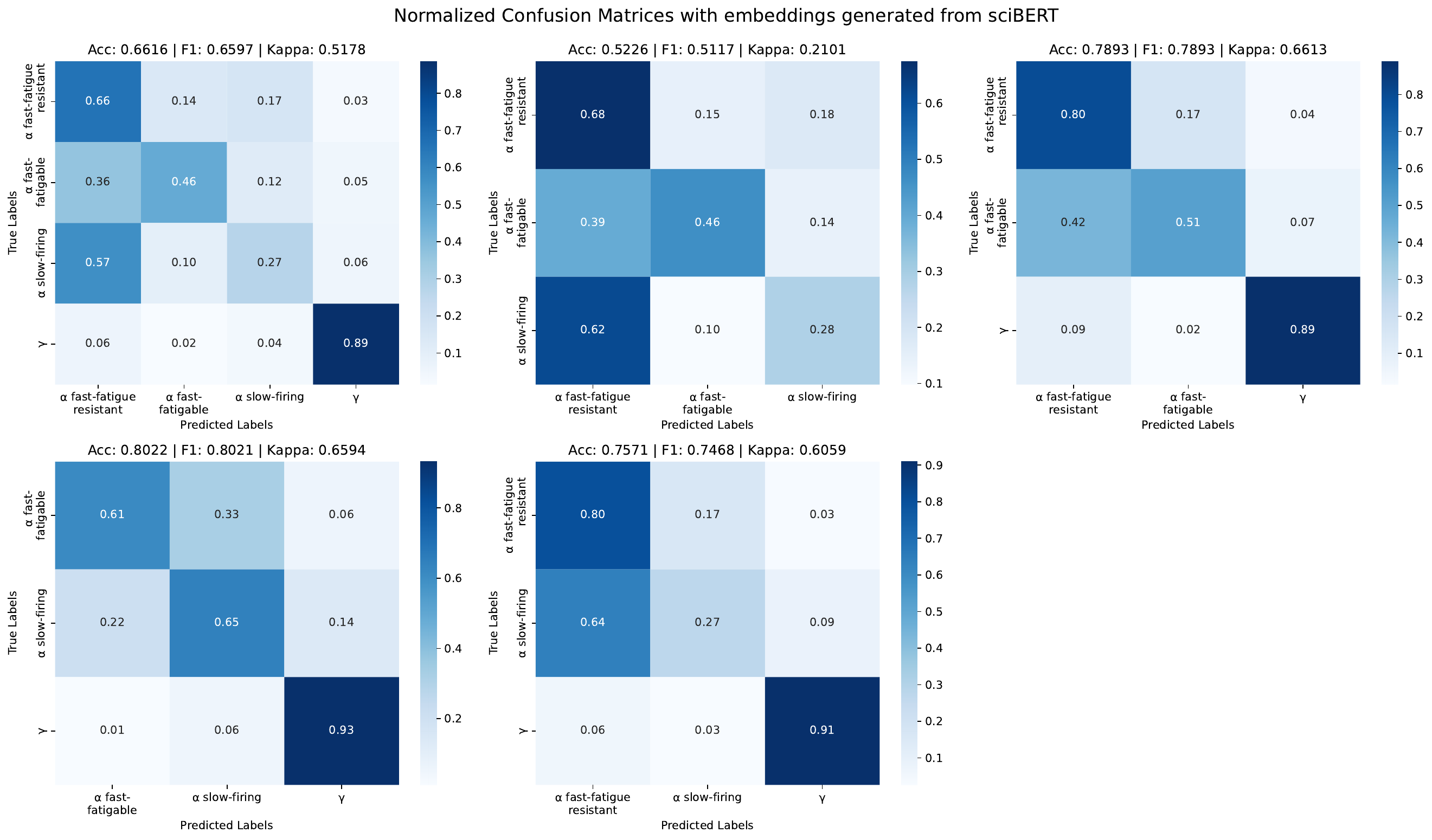}
  \caption{Confusion matrix using SciBERT embeddings.}
  \label{fig:cm_scibert}
\end{figure*}

To assess the sensitivity of the KNN classifier to the choice of neighborhood size \(k\), we conducted an ablation study using embeddings generated by the text-embedding-ada-002 model and classified the MNs using cosine distance. The classification task involved distinguishing four subtypes of MNs based on their expression-derived embeddings. As shown in Table~\ref{tab:knn_ablation}, performance improved consistently from \(k = 1\) to \(k = 50\) across all three metrics: accuracy, weighted F1 score, and Cohen’s kappa. Notably, the classifier achieved its highest accuracy of \textbf{71\%} and kappa of \textbf{0.57} at \(k = 50\), suggesting that moderate neighborhood sizes enable the model to robustly aggregate contextual similarity without overfitting to local noise. In contrast, very small \(k\) values (e.g., \(k=1\)) resulted in significantly lower accuracy (60\%) and kappa (0.44), indicating high variance and sensitivity to noisy neighbors. As \(k\) increased beyond 50, all metrics declined, likely due to over-smoothing, where large neighborhoods incorporate dissimilar cell types and blur class boundaries. These results highlight the importance of tuning \(k\) to balance bias and variance in embedding-based classification.

\begin{table}[htbp]
\centering
\caption{Ablation study evaluating KNN performance with varying neighborhood sizes \(k\), using embeddings generated by the text-embedding-ada-002 model and cosine distance}
\label{tab:knn_ablation}
\begin{tabular}{c|c|c|c}
\hline
\textbf{k} & \textbf{Accuracy} & \textbf{F1 Score (weighted)} & \textbf{Cohen's Kappa} \\
\hline
1   & 0.60 & 0.61 & 0.44 \\
3   & 0.64 & 0.64 & 0.49 \\
5   & 0.66 & 0.65 & 0.52 \\
10  & 0.69 & \textbf{0.68} & 0.56 \\
25  & 0.71 & 0.68 & 0.57 \\
50  & \textbf{0.71} & 0.6692 & \textbf{0.57} \\
100 & 0.69 & 0.63 & 0.53 \\
500 & 0.65 & 0.55 & 0.47 \\
\hline
\end{tabular}
\end{table}

\begin{figure}[htbp]
  \centering
  \includegraphics[width=\linewidth]{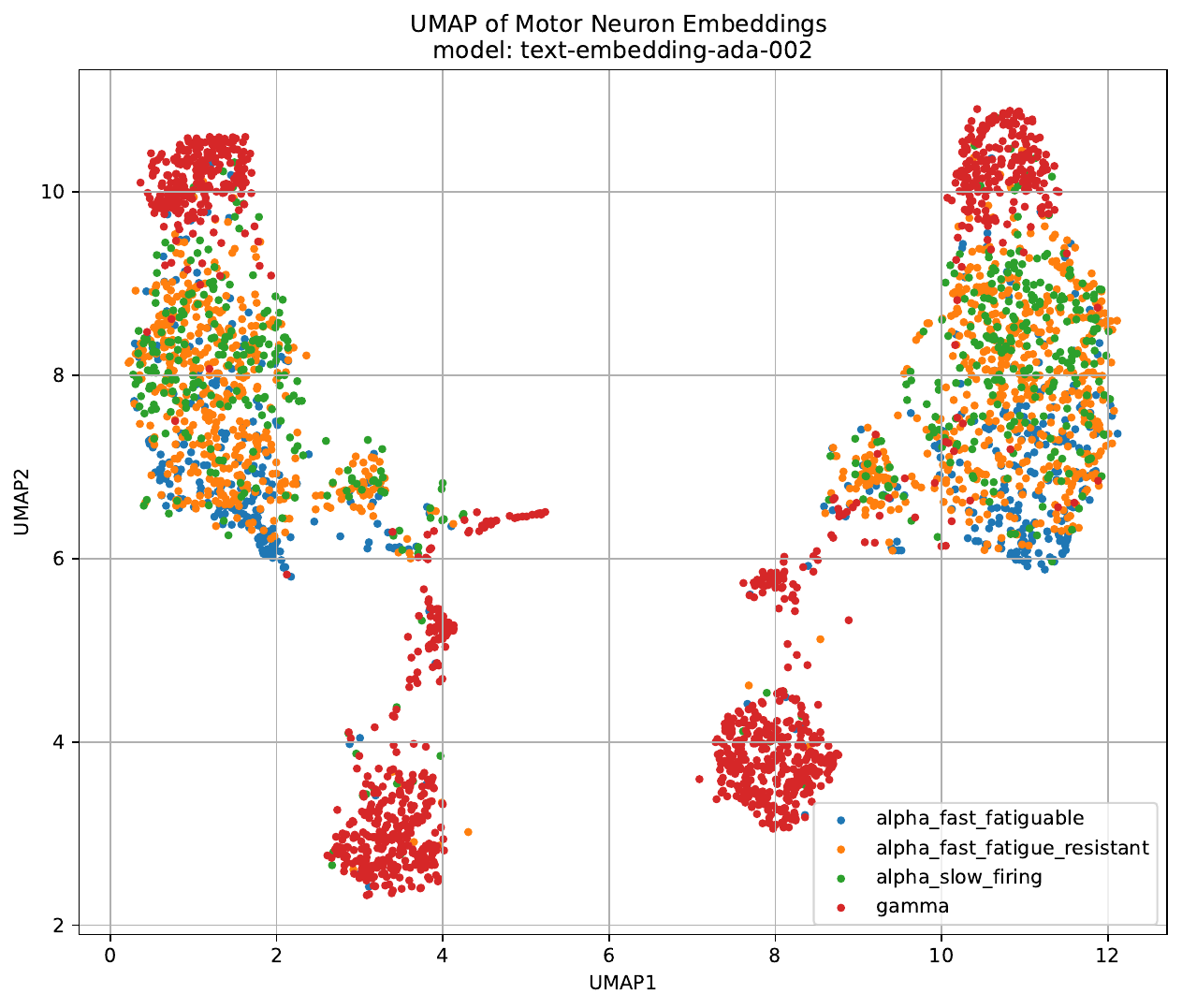}
  \caption{UMAP projection using text-embedding-ada-002 embeddings.}
  \label{fig:umap_ada002}
\end{figure}

\begin{figure}[htbp]
  \centering
  \includegraphics[width=\linewidth]{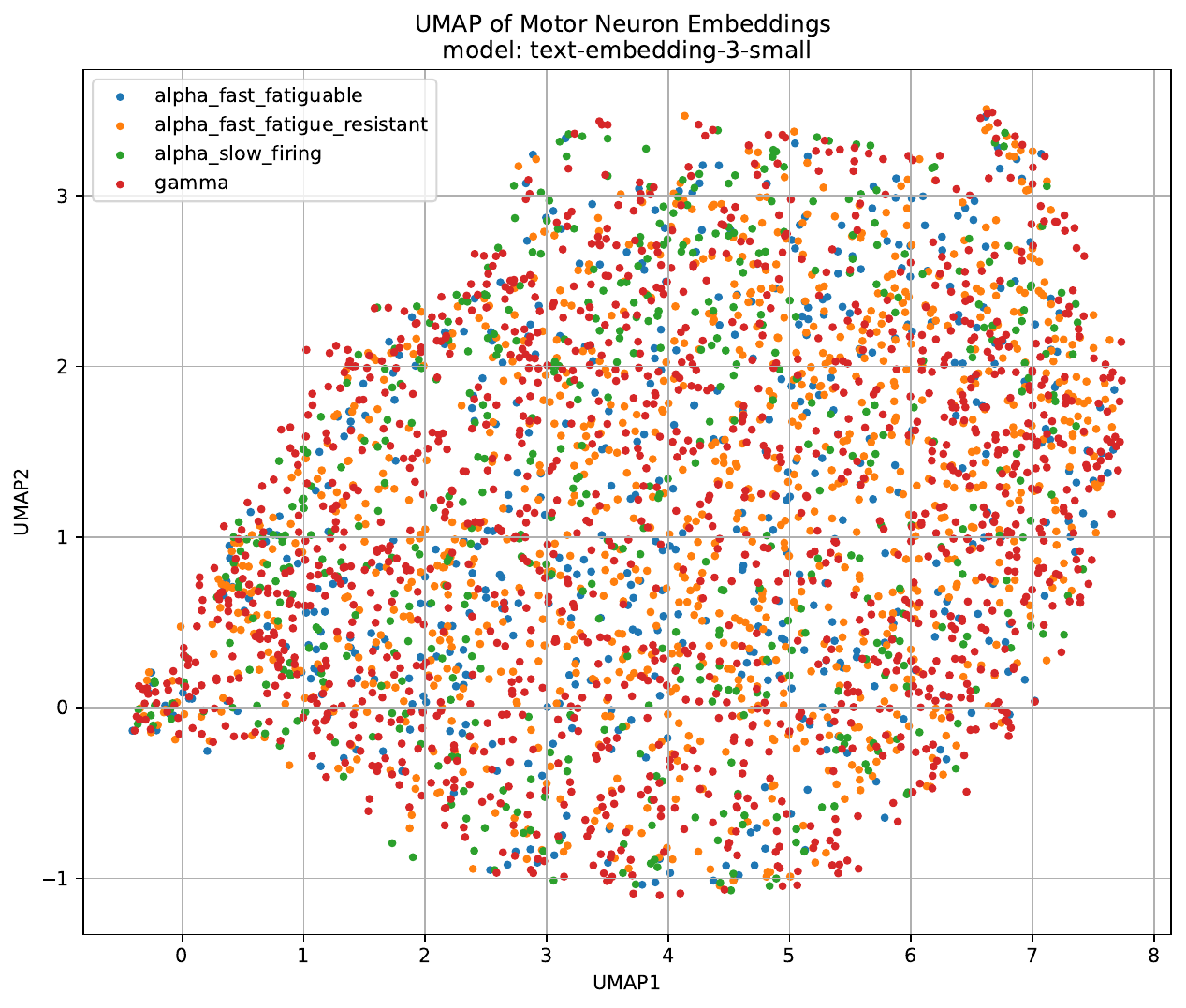}
  \caption{UMAP projection using text-embedding-3-small embeddings.}
  \label{fig:umap_3small}
\end{figure}

\begin{figure}[htbp]
  \centering
  \includegraphics[width=\linewidth]{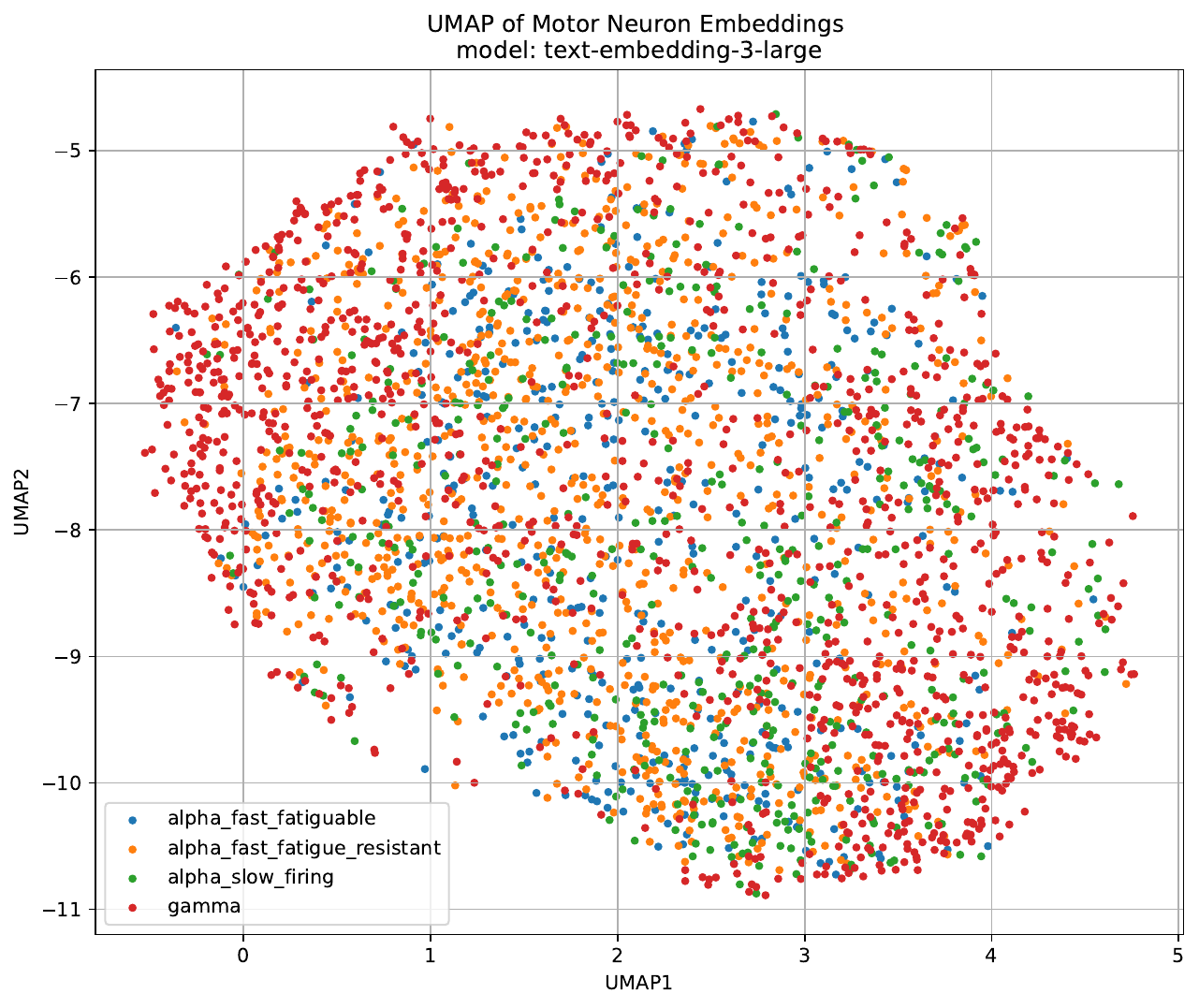}
  \caption{UMAP projection using text-embedding-3-large embeddings.}
  \label{fig:umap_3large}
\end{figure}

\begin{figure}[htbp]
  \centering
  \includegraphics[width=\linewidth]{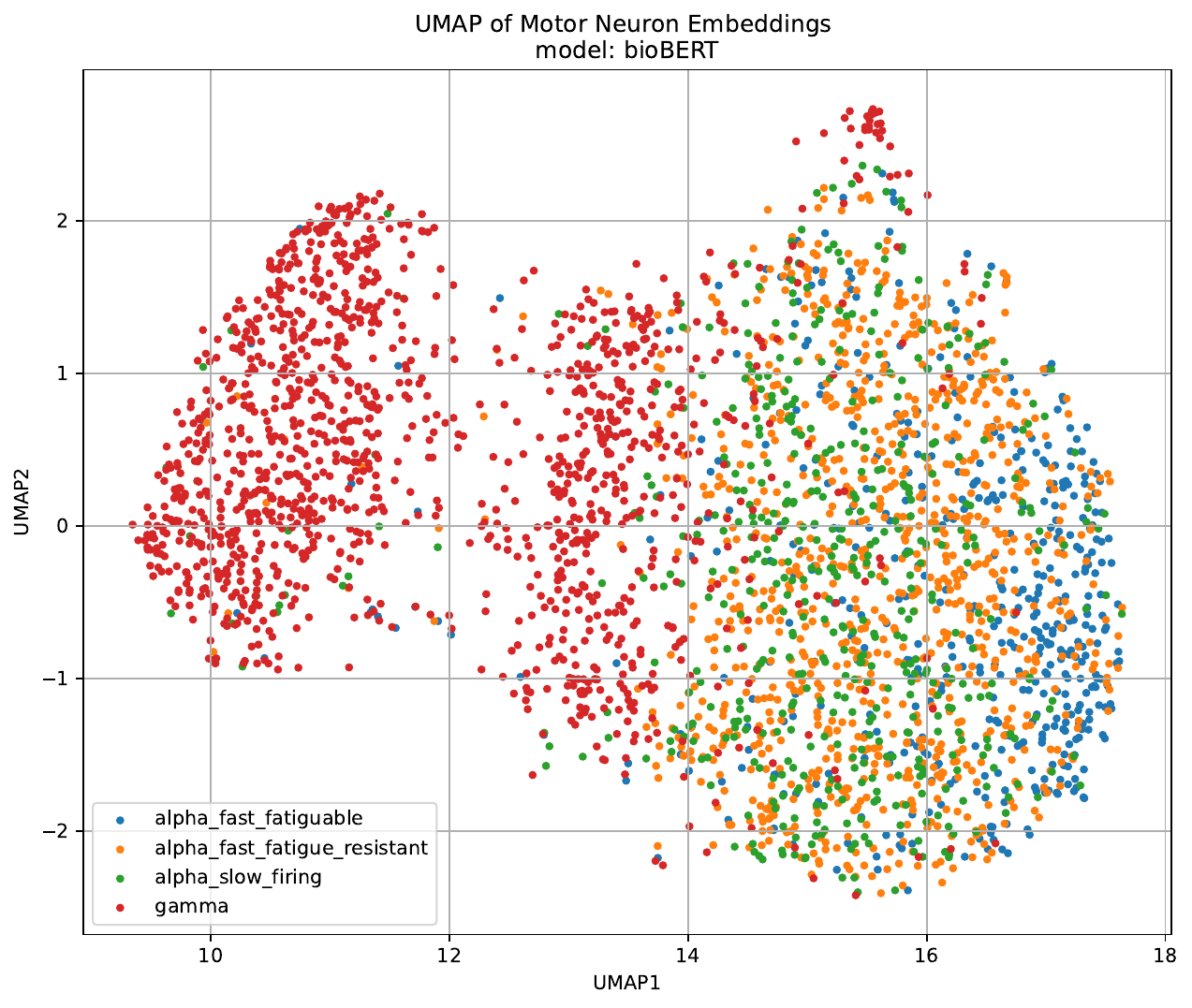}
  \caption{UMAP projection using BioBERT embeddings.}
  \label{fig:umap_biobert}
\end{figure}

\begin{figure}[htbp]
  \centering
  \includegraphics[width=\linewidth]{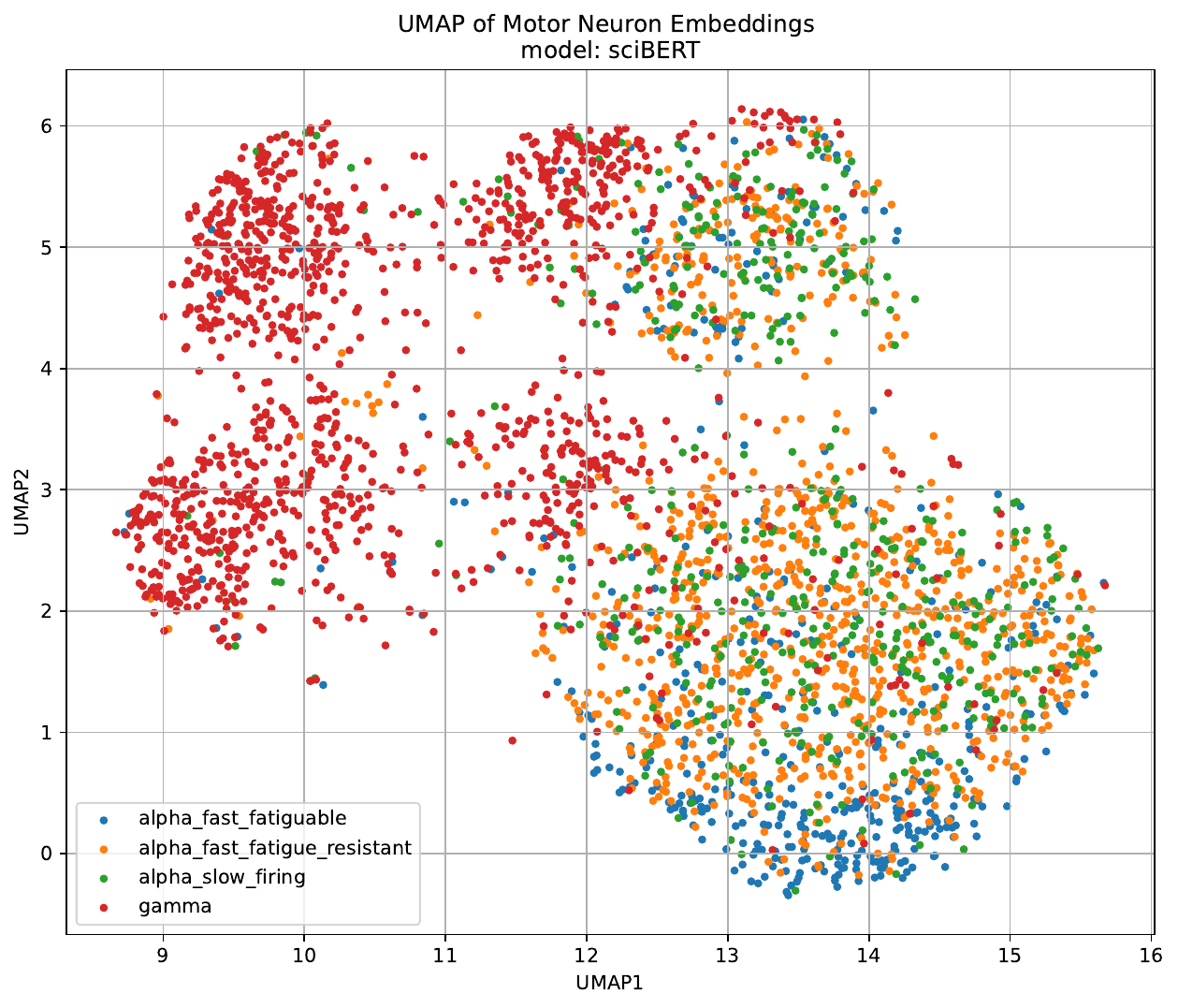}
  \caption{UMAP projection using SciBERT embeddings.}
  \label{fig:umap_scibert}
\end{figure}

To qualitatively assess the clustering capacity of different embedding models, we visualized the MN populations using UMAP. As shown in Figures~\ref{fig:umap_ada002} to~\ref{fig:umap_scibert}, the resulting projections varied significantly in their ability to preserve subtype separation. Among the embedding-based approaches, text-embedding-ada-002 and SciBERT exhibited the most distinct clustering, with partially separated clusters corresponding to the four biological subtypes: $\alpha$-FF, $\alpha$-FR, $\alpha$-SF, and $\gamma$ MNs. These visual impressions are supported by clustering metrics such as the Silhouette score, Calinski-Harabasz index, and Davies-Bouldin score (Table~\ref{tab:UMAP performance}), where ada-002 achieved the highest Calinski-Harabasz score (62.22), SciBERT showed strong overall balance, and BioBERT attained the highest Silhouette score (0.0018).

In contrast, text-embedding-3-small and text-embedding-3-large produced less structured UMAPs, with substantial subtype overlap and lower clustering metric scores, indicating reduced biological interpretability. These findings emphasize that the ability of embedding models to preserve fine-grained transcriptomic structure varies substantially and that domain-specific models such as BioBERT and SciBERT are generally more effective than general-purpose alternatives. Nevertheless, even the strongest LLM-based embeddings exhibit limitations in separating transcriptionally similar cell subtypes, highlighting the ongoing need for task-aware embedding strategies in single-cell analysis.

\begin{table}[ht]
\centering
\caption{Comparison of UMAP performances for motor neuron clustering}
\label{tab:UMAP performance}
\resizebox{\linewidth}{!}{%
\begin{tabular}{lccc}
\hline
\textbf{Method} & \textbf{Silhouette} & \textbf{Calinski-Harabasz} & \textbf{Davies-Bouldin} \\
\hline
text-embedding-ada-002 & -0.0027 & \textbf{62.22} & \textbf{9.76} \\
text-embedding-3-small & -0.0093 & 11.65 & 16.14 \\
text-embedding-3-large & -0.0047 & 16.28 & 13.38 \\
BioBERT & \textbf{0.0018} & 48.40 & 14.01 \\
SciBERT & 0.0007 & 50.83 & 11.79 \\
\hline
\end{tabular}
}
\end{table}

\section{Discussion}

We evaluated the utility of LLM-derived gene embeddings for classifying MN subtypes and their vulnerability using scRNA-seq/snRNA-seq dataset. Our approach transformed biologically relevant NCBI gene descriptions into dense vector representations using five pre-trained language models: three general-purpose models (text-embedding-ada-002, text-embedding-3-small, and text-embedding-3-large) and two domain-specific biomedical models (BioBERT and SciBERT). Cell-level embeddings were computed by taking expression-weighted averages of the top-expressed genes, allowing for the integration of semantic gene annotations with quantitative expression data.

Among the evaluated models, text-embedding-ada-002 achieved the highest classification performance, followed closely by BioBERT and SciBERT. In contrast, text-embedding-3-small and text-embedding-3-large showed notably lower performance, suggesting that domain relevance and model architecture play critical roles in the effectiveness of LLM-derived embeddings for biological tasks, such as gene expression profile analyses. These results indicate that while LLM-based embeddings can effectively recapitulate broad biological semantics, including cell identity or vulnerability to disease or degeneration. Nevertheless, their performance declines when tasked with resolving the fine-scale transcriptomic heterogeneity essential for precise classification of fine-grained MN subtypes (such as $\alpha$-FR MNs, $\alpha$-SF MNs, and $\alpha$-FF MNs). This limitation may be overcome by increasing the number of cells sampled for each fine-grained subtype. However, this approach is not always biologically feasible and may substantially increase the complexity and cost of scRNA-seq/snRNA-seq experiments.

From a computational perspective, the performance differences between these models probably stem from the fundamental distinctions of how each model generate embeddings. The OpenAI embedding models: text-embedding-ada-002, text-embedding-3-small and text-embedding-3-large are transformer-based architectures pre-trained on broad-domain corpora that include websites, books, and code repositories. These models are trained using contrastive learning objectives, which optimize the cosine similarity between embeddings of semantically similar inputs while increasing the distance between unrelated pairs. As a result, they produce fixed-size, sentence-level embeddings $\mathbf{z} \in \mathbb{R}^d$ that encode global semantic relationships across the input sequence. Notably, text-embedding-ada-002 is optimized for semantic search and retrieval tasks, which likely enhances its ability to cluster gene descriptions with shared biological functions in latent space. While text-embedding-3-small and text-embedding-3-large are more recent models designed for efficiency and general-purpose semantic performance, they may be less tuned for capturing the domain-specific subtleties required for transcriptomic classification.

In contrast, BioBERT and SciBERT are based on the BERT encoder architecture, which generates contextualized token-level embeddings $\mathbf{h}_i \in \mathbb{R}^d$ for each input token. These models are pre-trained using masked language modeling objectives on domain-specific corpora: PubMed and PMC full-text articles for BioBERT, and a multidisciplinary collection of scientific texts for SciBERT. To obtain a sentence-level representation, a pooling operation must be applied to the output token embeddings. Unlike the OpenAI models, these BERT-based embeddings are not explicitly optimized for semantic alignment at the sentence level, which may limit their ability to separate functionally distinct gene descriptions in downstream tasks. BioBERT's exclusive biomedical focus provides greater sensitivity to gene-specific terminology compared to SciBERT, potentially explaining its slightly higher classification performance in our results.

Additional analyses revealed that classification accuracy was strongly influenced by the inclusion of $\gamma$ MNs, which exhibited clearer transcriptomic separability compared to the more transcriptionally similar $\alpha$ subtypes. This observation highlights the importance of subtype-specific gene expression profiles and their impact on model performance. Not surprisingly, while our models were effective in distinguishing $\gamma$ and $\alpha$ MNs, accurately separating the various $\alpha$ MN subtypes proved to be much more challenging. These subtypes share more subtle transcriptomic differences, making them harder to resolve even with sophisticated embedding strategies. This limitation underscores the inherent biological complexity of the $\alpha$ MN subtypes, which certainly exist in most of the neuronal and non-neuronal cell subtypes, when a biologically relevant fine classification is considered. To tackle this challenge from a computational perspective, future work could explore contrastive learning approaches to better emphasize the subtle differences between subtypes, use gene-level attention mechanisms to focus on the most informative genes, or adopt hierarchical classification models that gradually refine broad predictions into more detailed subtype categories~\cite{pmlr-v119-chen20j, Li2025-rc}.

Our framework can be further extended by incorporating spatial transcriptomics or temporal dynamics to enhance the granularity of cell subtype classification. Spatial transcriptomics would enable the mapping of transcriptomic profiles back to anatomical context, allowing for the identification of spatially localized expression patterns that are not readily apparent in conventional scRNA-seq/snRNA-seq data with tissue dissociation~\cite{Zeng2023-fs, Rodriques2019-vn}. This integration could be especially valuable for resolving functionally distinct subpopulations within morphologically or transcriptionally similar groups. Similarly, the embedding temporal information, such as developmental stages, disease progression, or cellular response trajectories, could help disentangle subtype transitions and plasticity over time. These additions would not only refine classification accuracy but also provide biologically meaningful insights into how MN subtypes evolve and interact in their native tissue environments.

\section{Conclusion}

Our findings illustrate both the promise and the current limitations of applying pre-trained LLM embeddings in single-cell transcriptomic analysis. While embedding-based approaches provide a scalable and interpretable framework for incorporating biological text annotations, future work may benefit from hybrid strategies that combine semantic information with raw expression values to more accurately model subtle gene expression patterns in complex cell populations.

Beyond classifying MN subtypes or disease vulnerability, our framework is readily accessible for cross-cell-type comparisons among all neuronal and non-neuronal cell types. For example, future studies can directly extend our analytic pipeline to integrate multiple neuronal populations, such as MNs, retinal ganglion cells (RGCs), or hippocampal neurons—, to identify shared or divergent transcriptomic and functional properties~\cite{Tran2019-gd, Bugeon2022-rf}. This comparative strategy may yield new insights into cell-type-specific vulnerabilities in neurodegenerative conditions, such as ALS/MND, glaucoma and AD. Specifically, given the recent deposition of multiple disease-relevant scRNA-seq and snRNA-seq datasets in NCBI, our embedding-based methodology has the potential to be generalized across these different types of diseases. By applying the same or a fine-tuned version of our framework developed here to these datasets, conserved molecular signatures of neurodegeneration can be identified, either specific to individual diseases or shared across related conditions. In addition, early disease biomarkers or stratify disease subtypes can also be defined based on the single-cell high-resolution transcriptomic profiles~\cite{Zhou2020-nl}. The interpretability of our approach—driven by biologically grounded gene annotations—offers a valuable tool for investigating complex disease mechanisms across diverse neuronal systems.

Additionally, refining the embedding generation process may further improve classification performance. For instance, applying Z-score normalization prior to computing expression-weighted embeddings could reduce biases introduced by high-abundance genes and better highlight differentially expressed features. Incorporating cell-specific context, spatial transcriptomics, or temporal trajectories could also enhance the biological resolution of embeddings.

Overall, our study lays a foundation for applying LLM-based embeddings in single-cell analysis and provides a flexible framework that future computational and experimental efforts can build upon to better understand the transcriptional architecture and pathological mechanisms underlying neurodegeneration.


%

\section*{Acknowledgment}

This work was supported by grants from the National Institutes of Health (EY032181), ALS Association Seed Grant, Hop On A Cure Foundation Grant, and Brightfocus Alzheimer's Disease Research Grant. We thank Dr.~Aaron D. Gitler and colleagues for generating and making publicly available the snRNA-seq dataset \textbf{GSE161621}, which provided the foundational data for our analysis. Their work has significantly contributed to our understanding of motor neuron subtypes and their differential vulnerability in the adult mouse spinal cord.

\section*{Conflict of Interest}
F.~T. and D.~J. are co-founders of Regenerative AI LLC. The remaining authors declare no conflict of interest.

\ifCLASSOPTIONcaptionsoff
  \newpage
\fi



%
\vspace{1cm}
\bibliographystyle{IEEEtran}
\bibliography{references.bib}

\end{document}